\documentclass{article}
\usepackage[utf8]{inputenc}
\usepackage[english]{babel}
\usepackage{amssymb,amsmath}
\usepackage{xcolor}
\usepackage{tikz}
\usepackage{hyperref}

\usepackage{amsthm}
\theoremstyle{definition}
\newtheorem{example}{Example}[section]
\theoremstyle{remark}

\usetikzlibrary{positioning,fit,calc,petri}

\title{
    Algorithms and data structures for automatic precision estimation of neural networks
}
\author{
    Igor V. Netay
    \thanks{Joint Stock "Research and production company ``Kryptonite"}
    \thanks{Institute for Information Transmission Problems, Russian Academy of Sciences}
    \href{mailto:i.netay@kryptonite.ru}{i.netay@kryptonite.ru}
}
\date{}

\begin{document}
\maketitle

\begin{abstract}
    We describe algorithms and data structures to extend a neural network 
    library with automatic precision estimation for floating point computations.
    We also discuss conditions to make estimations exact and preserve high
    computation performance of neural networks training and inference.
    Numerical experiments show the consequences of significant precision loss for particular
    values such as inference, gradients and deviations from mathematically predicted behavior.

    It turns out that almost any neural network accumulates computational inaccuracies.
    As a result, its behavior does not coincide with predicted by the mathematical model of neural network.
    This shows that tracking of computational inaccuracies is important for
    reliability of inference, training and interpretability of results.
\end{abstract}

\section*{Introduction}

In modern industry, neural networks are widely used.
This is due to their great performance and highly developed mathematical background.
Theoretical results start with famous result of Kolmogorov and 
Arnold~\cite{Arnold2009,Arnold2009_2} disproving ``13th Hilbert problem''.
For general activations, there is also famous Cybekno approximation 
theorem~\cite{10.1007/978-1-4612-2856-1_21,Cybenko1989ApproximationBS,Funahashi1989OnTA}.
There are also many approximation results on many types of neural networks 
(\cite{Kidger2019UniversalAW,Geuchen2023UniversalAW,baker1998universal,kratsios2022universal,voigtlaender2023universal,qi2025universal,nguyen2016universal,gonon2025universal,yu2021arbitrary,Sonoda2024ConstructiveUA}).
All these positive results are based on real-valued or complex-valued mathematical models of neural network.

More recent papers are focused on negative approximation 
results~\cite{song2017complexity,kon2000information}.
Although many lower bounds can be proved for real-valued mathematical models of neural networks,
actually only computational models are applied.
They deal with $32$-bit, $16$-bit or sometimes more quantized models.
Many more recent papers~\cite{liu2024kan,somvanshi2024survey,ji2024comprehensive,vaca2024kolmogorov,kiamari2024gkan,bodner2024convolutional,aghaei2024rkan,cheon2024demonstrating,bresson2024kagnns,koenig2024kan,shen2025reduced} are still inspired by mathematical results for real-valued models.
Actually, there are only few positive or semi-positive approximation results for floating point 
numbers (like $32$-bit) instead of real-valued numbers~\cite{hwang2025floating}.
There are some negative results (see~\cite{vietrov2024methodological}).
Furthermore, positive results can be rarely possible in the framework 
of mainly indeterministic computational model depending on parallelization methods.

Computations using floating point numbers lead to inaccuracies.
These inaccuracies can accumulate.
Loss of significance for neural networks can lead to unreliable inference, training or results interpretation.
Goal of this paper is to provide methods to estimate inaccuracies and automagically
track them in neural networks inference and backpropagation.

We call a \textit{library} a collection of data structures, algorithms, methods
and their numerical implementations.
We will consider two libraries:
\begin{itemize}
    \item \textit{basic library} for neural networks (actually, it will be \texttt{torch} library),
    \item \textit{extending library} with automatic precision estimation.
        We call our extension ``\texttt{xtorch}''.
\end{itemize}
Although our extension library is based on \texttt{torch} library, we formalize
conditions on a basic library to provide it with such extension.
So, if a basic library satisfies these conditions, it can be potentially extensible in the same way.

When using the basic library, some tensors are generated as a result.
If we use the extending library, we get pairs of floating point tensor and \texttt{uint8} tensor
containing number of exact mantissa bits for the elements of the first one.
We impose some restrictions on implementation of~\texttt{xtorch}:
\begin{itemize}
    \item When it is possible, values calculated with basic and extending libraries should coincide.
    \item When it is possible, precision estimations should be based on exact computations.
    \item Implementation is based on implicit backpropagation algorithm with
        garbage collection. So, we should ensure that memory used by extending 
        library can be freed by the garbage collector also.
\end{itemize}

\section{Results}

Library \texttt{xtorch} is a \texttt{Python} library extending \texttt{torch}
with classes, methods and functions for neural networks with automatic
precision estimation for \texttt{float32} and \texttt{float64}.

In the case of computations on CPU, \texttt{xtorch} is based on \texttt{XNumPy}~(see~\cite{xnumpy_repo}) library.
For computations on GPU, \texttt{xtorch} provides own implementations.

% Coverage of methods ???

The main results of this paper are the following:
\begin{itemize}
    \item Assumptions on a neural network framework to
        extend it with automatic precision estimation are formalized and listed.
    \item Algorithms, data structures and methods for
        automatic precision estimation were designed and implemented.
    \item Numerical experiments were conducted, consequences of inaccuracies accumulation are described.
    \item Hypothesis on correlation of loss function fluctuation 
        and accumulation of numerical inaccuracies was confirmed.
\end{itemize}

\section{Related work}

Optimization of neural networks for floating point computations is mainly 
focused on quantization methods and performance 
improvement~\cite{romanov2021analysis,agrawal2019dlfloat,trusov2025training,guo2025pt}.
At the same time the more network is quantized the less numerically stable results are.
Evaluation of functions with inexact arguments leads to inexact results.
The dependence of inexactness of the result on inexactness of arguments is
expressed by the \textit{condition number}.

The tensors of weights learned by networks providing mapping between layers, which usually have
high condition numbers~\cite{saarinen1993ill}.
Such maps are said to be \textit{ill-conditioned}.
This leads to high numerical errors and can even lead to different vulnerabilities~\cite{sinha2018neural,krishnamurthyneural}.
Ill-conditioning of neural networks demonstrates a hard training 
problem (see~\cite{van2002solving,cao2025analysis,burger2000training,li2005network}).
This problem reveals that some common approach of inaccuracy tracking for neural networks is needed.
It is the main goal of current paper.

\section{Assumptions on a basic neural network library}

\begin{figure}[ht]
\centering
\begin{tikzpicture}
    \node (t) {DAG node};
    \matrix (m) [draw,column sep={3mm,between origins}, nodes=draw, below = 1mm of t]{
        \node (v) {value}; & \node [right = 1mm of v] {grad}; \\
    };
    \node [draw=black!50, fit={(t) (m)}] {};
\end{tikzpicture}
\caption{DAG node consists of values and gradient being floating point tensors.}
\label{lab:dag-node}
\end{figure}
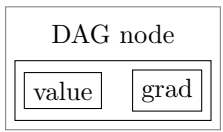

Let us give some basic notation to formalize neural network library structure.
We deal with computations as direct acyclic graphs (DAG).
Vertices of DAG correspond to computational operations.
They can contain some parameter tensors inside and have some placeholders to
substitute with the input tensors and proceed the computation.
For instance, linear layer contains weight and bias in every vertex.
During the inference the parameter tensors are not changed.
During training, they can change.
Also, gradients are included into DAG vertices (gradient is either tensor or
an absent value corresponding to \texttt{None} object in \texttt{Python}).
Contents of DAG node is shown on~Fig.\,\ref{lab:dag-node}.

Neural networks presuppose backpropagation.
This makes necessary to store DAG and some intermediate tensors.
All the nodes in the computational graph contain some auto differentiable function.

The \textit{neural network model} is data structure consisting of
\begin{itemize}
    \item a DAG $\mathfrak{N}$ with a partial ordering $<$ and lists
        $\mathfrak{n}_{\text{in}}$ and $\mathfrak{n}_{\text{out}}$ of input
        and output nodes for each node~$\mathfrak{n}\in\mathfrak{N}$;
    \item a forward function $\mathfrak{f}_{\mathfrak{n}}$ consuming 
        $\sharp \mathfrak{n}_{\text{in}}$ floating point tensor arguments
        from incoming nodes and producing a floating point tensor for each $\mathfrak{n}\in\mathfrak{N}$;
    \item a backward function $\mathfrak{b}_{\mathfrak{n}}$ for each 
        $\mathfrak{n}\in\mathfrak{N}$ consuming a tensor produced 
        by~$\mathfrak{f}_{\mathfrak{n}}$, and tensors produced by 
        $\{\mathfrak{f}_\mathfrak{k}\}_{\mathfrak{k}\in\mathfrak{n}_{\text{out}}}$ and the sum of all
        $\{\mathfrak{b}_\mathfrak{k}\}_{\mathfrak{k}\in\mathfrak{n}_{\text{out}}}$.
\end{itemize}

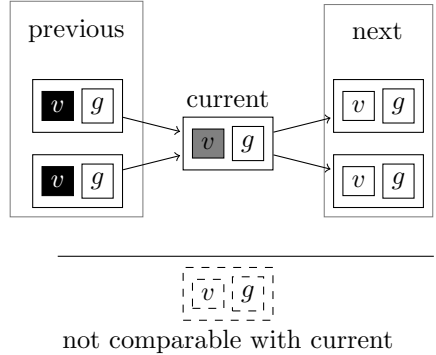
\begin{figure}[ht]
\centering
\begin{tikzpicture}[node distance=3mm]
    \node (t) at (0.25,1) {previous};

    \node (a1) at (0,0) [draw, fill=black!50, black] {\color{white}{$v$}};
    \node (a2) [draw, right = 1mm of a1] {$g$};
    \node (a) [draw, fit={(a1) (a2)}] {};

    \node (b1) at (0,-1) [draw, fill=black] {\color{white}{$v$}};
    \node (b2) [draw, right = 1mm of b1] {$g$};
    \node (b) [draw, fit={(b1) (b2)}] {};

    \node (c) [draw=black!50, fit={(t) (a) (b)}] {};

    \node (d1) at (2,-0.5) [draw, fill=black!50!white] {$v$};
    \node (d2) [draw, right = 1mm of d1] {$g$};

    \node (s) at (4.25,1) {next};

    \node (e1) at (4,0) [draw] {$v$};
    \node (e2) [draw, right = 1mm of e1] {$g$};
    \node (e) [draw, fit={(e1) (e2)}] {};

    \node (f1) at (4,-1) [draw] {$v$};
    \node (f2) [draw, right = 1mm of f1] {$g$};
    \node (f) [draw, fit={(f1) (f2)}] {};

    \node (g) [draw=black!50, fit={(s) (e) (f)}] {};

    \node (d) [draw, label=90:current, fit={(d1) (d2)}] {}
        edge [pre] (a)
        edge [pre] (b)
        edge [post] (e)
        edge [post] (f)
    ;

    \node (h1) at (2,-2.5) [draw, dashed] {$v$};
    \node (h2) [draw, dashed, right = 1mm of h1] {$g$};
    \node (h) [draw, dashed, fit={(h1) (h2)}, label=270:not comparable with current] {};

    \draw (0cm,-2cm) -- (5cm,-2cm);

\end{tikzpicture}
\caption{Forward. Already computed tensors are black, current node is grey, not computed tensors are white}
\label{lab:dag-fwd}
\end{figure}

\begin{figure}[ht]
\centering
\begin{tikzpicture}[node distance=3mm]
    \node (t) at (0.25,1) {previous};

    \node (a1) at (0,0) [draw, fill=black!50, black] {\color{white}{$v$}};
    \node (a2) [draw, right = 1mm of a1] {$g$};
    \node (a) [draw, fit={(a1) (a2)}] {};

    \node (b1) at (0,-1) [draw, fill=black] {\color{white}{$v$}};
    \node (b2) [draw, right = 1mm of b1] {$g$};
    \node (b) [draw, fit={(b1) (b2)}] {};

    \node (c) [draw=black!50, fit={(t) (a) (b)}] {};

    \node (d1) at (2,-0.5) [draw, fill=black] {\color{white}{$v$}};
    \node (d2) [draw, fill=black!50!white, right = 1mm of d1] {$g$};

    \node (s) at (4.25,1) {next};

    \node (e1) at (4,0) [draw, fill=black] {\color{white}{$v$}};
    \node (e2) [draw, fill=black, right = 1mm of e1] {\color{white}{$g$}};
    \node (e) [draw, fit={(e1) (e2)}] {};

    \node (f1) at (4,-1) [draw, fill=black] {\color{white}{$v$}};
    \node (f2) [draw, fill=black, right = 1mm of f1] {\color{white}{$g$}};
    \node (f) [draw, fit={(f1) (f2)}] {};

    \node (g) [draw=black!50, fit={(s) (e) (f)}] {};

    \node (d) [draw, label=90:current, fit={(d1) (d2)}] {}
        edge [post] (a)
        edge [post] (b)
        edge [pre] (e)
        edge [pre] (f)
    ;

    \node (h1) at (2,-2.5) [draw, dashed] {$v$};
    \node (h2) [draw, dashed, right = 1mm of h1] {$g$};
    \node (h) [draw, dashed, fit={(h1) (h2)}, label=270:not comparable with current] {};

    \draw (0cm,-2cm) -- (5cm,-2cm);

\end{tikzpicture}
\caption{Backward. Already computed tensors are black, current node is grey, not computed tensors are white}
\label{lab:dag-bwd}
\end{figure}
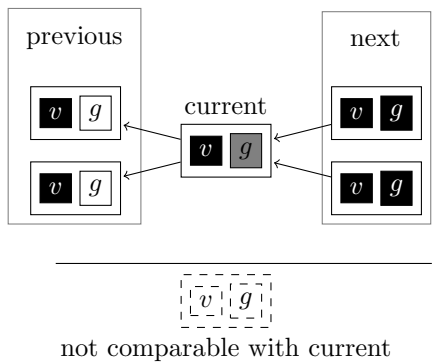

The order of computations assumes the following conditions:
\begin{itemize}
    \item when $\mathfrak{f}_{\mathfrak{n}}$ is being computed, all
        $\{\mathfrak{f}_{\mathfrak{k}}\}_{\mathfrak{k} < \mathfrak{n}}$ should have been
        already computed (see~Fig.\,\ref{lab:dag-fwd});
    \item when $\mathfrak{g}_{\mathfrak{n}}$ is being computed, all
        $\{\mathfrak{g}_{\mathfrak{k}}\}_{\mathfrak{k} > \mathfrak{n}}$
        and
        $\{\mathfrak{f}_{\mathfrak{k}}\}_{
            \mathfrak{k} > \mathfrak{n},
            \mathfrak{k} < \mathfrak{n},
            \mathfrak{k} = \mathfrak{n}
        }$
        should have been already computed (these nodes are comparable with $\mathfrak{n}$ 
        nodes, but not all the nodes, see~Fig.\,\ref{lab:dag-bwd}).
\end{itemize}

When~$\mathfrak{g}_\mathfrak{n}$ is being computed for some~$\mathfrak{n}\in\mathfrak{N}$,
all $\{\mathfrak{g}_{\mathfrak{k}}\}_{\mathfrak{k} > \mathfrak{n}}$ are computed, and their sum
is computed implicitly.
This detail will be important as we will show below.

For our further needs, we need some additional structure.
Let us call \textit{hook enhanced neural network model} a neural network model
with additional set of callable objects $\mathfrak{H}_\mathfrak{n}$ for each~$\mathfrak{n}\in\mathfrak{N}$.
Basically, each of these callable objects $\mathfrak{h}\in\mathfrak{H}_\mathfrak{n}$
consumes and produces a tensor of shape equal to shape of~$\mathfrak{g}_{\mathfrak{n}}$, but
we will assume this object can use some already computed data.
These objects will not be pure functions for our further needs.
Here we rely on Python duck typing ability and impose some additional restriction.
Namely, the order of computation should be the following:
\begin{itemize}
    \item all $\mathfrak{b}_{>\mathfrak{n}}$,
    \item all $\mathfrak{H}_\mathfrak{n}$,
    \item any of $\mathfrak{b}_{<\mathfrak{n}}$ and $\mathfrak{h}_{< \mathfrak{n}}$.
\end{itemize}

For instance,
\begin{itemize}
    \item \texttt{torch} library supports hook enhanced neural network model;
    \item \texttt{tensorflow} library supports the same if it is enriched with
        \texttt{tensorflow-hooks} package.
\end{itemize}

\section{Data structures for automatic precision estimation}

\begin{figure}[ht]
\centering
\begin{tikzpicture}
    \node (t) {DAG node};
    \matrix (m) [draw,column sep={24mm,between origins}, nodes=draw, below = 1mm of t]{
        \node (v) {value}; & \node {grad}; \\
        \node (veb) {exact bits}; & \node {grad exact bits}; \\
    };
    \node [draw=black!50, fit={(t) (m)}] {};
\end{tikzpicture}
\caption{
    Extended DAG node consists of values and gradient being floating point 
    tensors and their exactness being u8 tensor.
}
\label{lab:edag-node}
\end{figure}
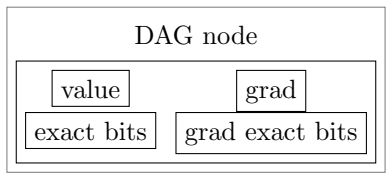

We extend nodes with \texttt{uint8} tensors for the number of mantissa exact bits corresponding
to tensor values and its gradient (correspondingly, \texttt{None} for gradient 
exact bits if gradient is represented by the \texttt{None} object, see~Fig.\,\ref{lab:edag-node}).

Basic library processes values and gradients.
Our goal is to make precision tensors automagically computed.
Note that all the computational graph processing is made implicitly by the basic library.
We piggy-back on this proceeding by structures defined in the extending library.
Precision of values is added by classes and function overloads in~\texttt{xtorch}.
Precision of gradients is computed by hooks (callable objects stored at tensors).

While precision of neural network inference is being computed, almost all
operations are called explicitly.
So, the precision estimation can be done by overloading of \texttt{Tensor}, \texttt{Module}
and some other classes and functions.
In \texttt{xtorch} basic classes are included as field in extending data structures 
(composition is applied instead of inheritance).
As it was done before in \texttt{XNumPy}, precision estimation for an atomic
computational operation is usually made either element-wise, or with
estimation of condition number.

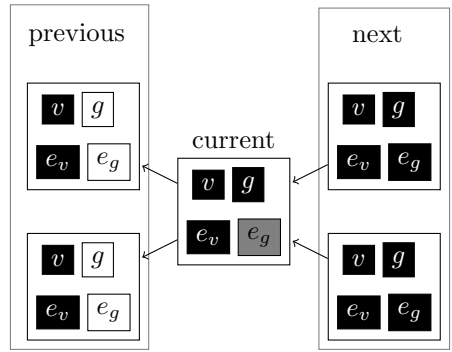
\begin{figure}[ht]
\centering
\begin{tikzpicture}[node distance=3mm]
    \node (t) at (0.25,1) {previous};

    \node (a11) at (0,0) [draw, fill=black!50, black] {\color{white}{$v$}};
    \node (a12) [draw, right = 1mm of a11] {$g$};
    \node (a21) at (0,-0.7) [draw, fill=black!50, black] {\color{white}{$e_v$}};
    \node (a22) [draw, right = 1mm of a21] {$e_g$};
    \node (a) [draw, fit={(a11) (a12) (a21) (a22)}] {};

    \node (b11) at (0,-2) [draw, fill=black] {\color{white}{$v$}};
    \node (b12) [draw, right = 1mm of b11] {$g$};
    \node (b21) at (0,-2.7) [draw, fill=black] {\color{white}{$e_v$}};
    \node (b22) [draw, right = 1mm of b21] {$e_g$};
    \node (b) [draw, fit={(b11) (b12) (b21) (b22)}] {};

    \node (c) [draw=black!50, fit={(t) (a) (b)}] {};

    \node (d11) at (2,-1.0) [draw, fill=black] {\color{white}{$v$}};
    \node (d12) [draw, fill=black, right = 1mm of d11] {\color{white}{$g$}};
    \node (d21) at (2,-1.7) [draw, fill=black] {\color{white}{$e_v$}};
    \node (d22) [draw, fill=black!50!white, right = 1mm of d21] {$e_g$};

    \node (s) at (4.25,1) {next};

    \node (e11) at (4,0) [draw, fill=black] {\color{white}{$v$}};
    \node (e12) [draw, fill=black, right = 1mm of e11] {\color{white}{$g$}};
    \node (e21) at (4,-0.7) [draw, fill=black] {\color{white}{$e_v$}};
    \node (e22) [draw, fill=black, right = 1mm of e21] {\color{white}{$e_g$}};
    \node (e) [draw, fit={(e11) (e12) (e21) (e22)}] {};

    \node (f11) at (4,-2) [draw, fill=black] {\color{white}{$v$}};
    \node (f12) [draw, fill=black, right = 1mm of f11] {\color{white}{$g$}};
    \node (f21) at (4,-2.7) [draw, fill=black] {\color{white}{$e_v$}};
    \node (f22) [draw, fill=black, right = 1mm of f21] {\color{white}{$e_g$}};
    \node (f) [draw, fit={(f11) (f12) (f21) (f22)}] {};

    \node (g) [draw=black!50, fit={(s) (e) (f)}] {};

    \node (d) [draw, label=90:current, fit={(d11) (d12) (d21) (d22)}] {}
        edge [post] (a)
        edge [post] (b)
        edge [pre] (e)
        edge [pre] (f)
    ;

\end{tikzpicture}
\caption{Backward. Already computed tensors are black, current node is grey, not computed tensors are white}
\label{lab:edag-bwd}
\end{figure}

While precision of gradients is being computed, all the computational operations
are called implicitly in correspondence with computational graph processing.
So, this computational graph <<does not know>> anything about extension library
and works independently without precision estimations.
Hook extension is the mechanism that do provide necessary computation facility
to be called and estimate precision of basic library computation (see~Fig.\,\ref{lab:edag-bwd}).

\vspace*{1cm}

\subsection{Conditions of gradients precision estimation consistence}

There are the following conditions:
\begin{itemize}
    \item Consider a vertex~$\mathfrak{n}$ in the computational graph.
        When we estimate precision of gradient at~$\mathfrak{n}$,
        all the values, gradients and their precision estimations should have been
        already computed for each node~$\mathfrak{k}$ if~$\mathfrak{k}>\mathfrak{n}$.
    \item Node~$\mathfrak{n}$ gets incoming gradients from computational
        operations where its produced value was involved.
        These incoming gradients should have been computed and summed already up at the
        beginning of precision estimation at~$\mathfrak{n}$.
    \item Produced by vertex~$\mathfrak{n}$ tensor is usually a temporary object.
        If this value is needed for precision estimation, a reference to it
        should be stored to prevent its elimination by garbage collector.
        At the same time, after precision estimation the reference should be deleted
        or replaced with a weak reference to make the temporary object garbage collectable.
\end{itemize}

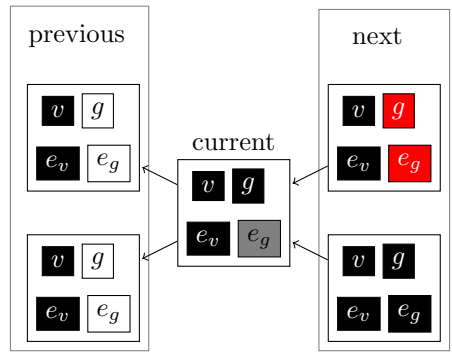
\begin{figure}[ht]
\centering
\begin{tikzpicture}[node distance=3mm]
    \node (t) at (0.25,1) {previous};

    \node (a11) at (0,0) [draw, fill=black!50, black] {\color{white}{$v$}};
    \node (a12) [draw, right = 1mm of a11] {$g$};
    \node (a21) at (0,-0.7) [draw, fill=black!50, black] {\color{white}{$e_v$}};
    \node (a22) [draw, right = 1mm of a21] {$e_g$};
    \node (a) [draw, fit={(a11) (a12) (a21) (a22)}] {};

    \node (b11) at (0,-2) [draw, fill=black] {\color{white}{$v$}};
    \node (b12) [draw, right = 1mm of b11] {$g$};
    \node (b21) at (0,-2.7) [draw, fill=black] {\color{white}{$e_v$}};
    \node (b22) [draw, right = 1mm of b21] {$e_g$};
    \node (b) [draw, fit={(b11) (b12) (b21) (b22)}] {};

    \node (c) [draw=black!50, fit={(t) (a) (b)}] {};

    \node (d11) at (2,-1.0) [draw, fill=black] {\color{white}{$v$}};
    \node (d12) [draw, fill=black, right = 1mm of d11] {\color{white}{$g$}};
    \node (d21) at (2,-1.7) [draw, fill=black] {\color{white}{$e_v$}};
    \node (d22) [draw, fill=black!50!white, right = 1mm of d21] {$e_g$};

    \node (s) at (4.25,1) {next};

    \node (e11) at (4,0) [draw, fill=black] {\color{white}{$v$}};
    \node (e12) [draw, fill=red, right = 1mm of e11] {\color{white}{$g$}};
    \node (e21) at (4,-0.7) [draw, fill=black] {\color{white}{$e_v$}};
    \node (e22) [draw, fill=red, right = 1mm of e21] {\color{white}{$e_g$}};
    \node (e) [draw, fit={(e11) (e12) (e21) (e22)}] {};

    \node (f11) at (4,-2) [draw, fill=black] {\color{white}{$v$}};
    \node (f12) [draw, fill=black, right = 1mm of f11] {\color{white}{$g$}};
    \node (f21) at (4,-2.7) [draw, fill=black] {\color{white}{$e_v$}};
    \node (f22) [draw, fill=black, right = 1mm of f21] {\color{white}{$e_g$}};
    \node (f) [draw, fit={(f11) (f12) (f21) (f22)}] {};

    \node (g) [draw=black!50, fit={(s) (e) (f)}] {};

    \node (d) [draw, label=90:current, fit={(d11) (d12) (d21) (d22)}] {}
        edge [post] (a)
        edge [post] (b)
        edge [pre] (e)
        edge [pre] (f)
    ;

\end{tikzpicture}
\caption{Backward. Already computed tensors are black, current node is grey, not computed tensors are white,
absent tensors are painted red.}
\label{lab:edag-bwd-ill}
\end{figure}

The condition above involves some additional backpropagation-time reference checking.
Consider the case where a hook stores an invalid reference to a tensor value 
and its number of exact bits (see~Fig.\,\ref{lab:edag-bwd-ill}).
It is a legal behavior.
It can occur if the temporary value was not involved in the loss function computation,
so it will not have a gradient.
At the same time, a hook stores a reference to a tensor without gradient.
In this case the hook call should be ignored, and the reference to temporary object should be freed or made weak.

Another memory leak possibility is when a hook stores a reference and is stored
in a tensor not being a temporary object.
So, if the hooks stores no data and only references, it creates a memory leak
if it is not deleted after the hook call.
So, hooks must be once-callable objects and must be freed after call.

These remarks are relied on the \texttt{Python} memory model based on reference counting.
A reference increments a referee reference counting, while weak reference does not.
So, a reference cannot be invalidated and a weak reference can.
Combination of strong and weak references provides a way to construct
complicated garbage collectable reference structures without manual memory management.

Although the hook mechanism was designed primarily to change or log gradients,
we utilize it to perform some additional computations with access to some
other data than the gradient itself.
Order of hook calls corresponds to order of their addition.
Actually, this does not matter for precision computation.
It is just needed to call them in some sequence without race condition.
We call hooks callable objects but not functions to point out that these objects
can store additional references or data and can access other data instead of the only gradient.
Luckily, the \texttt{Python} duck typing allows this hook usage.

We can conclude that if the precision estimation is performed in the way above, then
\begin{itemize}
    \item Precision estimation is guaranteed.
    \item Data access and precision estimation are consistent.
    \item Redundant references are freed and memory leaks are prevented.
\end{itemize}

\section{Atomic operations and precision estimation}

Computational graph construction and processing inside the basic library works
as a black box algorithm.
But the conditions listed above guarantee that precision estimation may be performed consistently.
Closed code implementation of the computational graph is not an obstacle for precision estimation.
At the same time, closed code implementation of a series of numeric algorithms
imposes some limitations for precision estimation.

\begin{example}
    Consider the operation of matrix multiplication.
    There are the following ways to estimate its precision:
    \begin{itemize}
        \item One can compute the product directly by the definition, count
            errors in arithmetic operations and find the difference of direct
            and fast results.
            This is the least computationally effective variant, although it can give the
            most accurate precision estimation.
        \item One can estimate the matrix product precision with a tropic matrix
            product estimation (see~\cite{xnumpy}).
            This way works fast, but it involves inner matrix multiplication
            and can lead to possibly inexact estimation if some matrix
            entries in the inner multiplication have no exact bits.
        \item Algorithm can be overloaded in such a way that each computational
            operations in its implementation would be provided with a precision
            estimation.
            In this way the estimation is deduced to estimation of arithmetic
            operations precision, so does not involve inexact operations, if
            the used compiler admits IEEE754 standard.
            At the same time, the source code of matrix multiplication is needed.
            It is available for BLAS and therefore matrix multiplication can
            be estimated for computations on CPU.
            At the same time, this is inapplicable for GPU matrix multiplication with \texttt{cuBLAS}.
    \end{itemize}

    We see that a computational operation with closed source code implementation 
    cannot be extended by precision estimation as a sequence of simpler 
    computational operations with accurate precision estimation.
    Instead, its precision can only be estimated as a precision of atomic computational operation.
    Also, the estimation can turn out to be inexact itself without very high loss of performance.

    It the case of computation on CPU third way is applied in \texttt{xtorch}.
    In the case of GPU the second one is applied.
\end{example}

\begin{example}
    Consider the linear layer computing just a linear function~$y=Ax+B$.

    \begin{itemize}
        \item Implementation deducing the computation to multiplication and
            then addition can lead to the results not equal to results obtained
            from \texttt{torch} library (multiplication and addition are made
            with one optimized function like \texttt{gemm} (BLAS) or \texttt{cuBlasGemm} (cuBLAS)).
        \item Some self-made implementation can also lead to different results,
            because implementation details (like parallelization) can give distinct results.
        \item Some indirect estimation like tropical product involves inner
            matrix multiplications and can lead to inaccuracies in the precision estimation.
    \end{itemize}

    In \texttt{xtorch} the first way is chosen for now.
    For CPU, it gives results the same as \texttt{torch} gives.
    For GPU, results and gradients will slightly differ.
\end{example}

These examples suggest the following general conclusion on precision estimation.
Suppose we were given an optimized numerical computation algorithm.
Then the following conditions can never hold all together:
\begin{itemize}
    \item the implementation has closed code,
    \item the precision estimation can be done without high performance loss,
    \item the precision estimation can be done without inexact inner computations,
        so it is mathematically guaranteed,
    \item new overloaded algorithm with precision estimation gives 
        identical results for any input data.
\end{itemize}

In other words, a closed source numerical algorithm cannot be provided with
mathematically exact precision estimation without algorithm change or performance loss.

For instance, matrix multiplication and other linear algebra algorithms from \texttt{BLAS} and \texttt{numpy}
can be provided with high performance mathematically correct precision estimations.
The same fails for \texttt{torch} and \texttt{cuBLAS}, so at least one 
of helpful estimation properties will be lost for any implementation.

There are many such computational operations like convolutions and Fast Fourier Transform.
Open implementation of numerical algorithms becomes a key part of 
mathematically reliable high-performance computation.
We will see below how computational inexactness can lead to non-reliable results.

\section{Numerical experiments and significance loss}

This section contains experiment results which were obtained with \texttt{xtorch-0.2.0} as extending library.
It is based on underlying \texttt{torch-1.13.1}, \texttt{xnumpy-1.0.2}, 
\texttt{numpy-1.24.4} for \texttt{Python3.8} over \texttt{Ubuntu22.04}.

\subsection{Fluctuations and accumulated inaccuracies}

Fluctuations of loss-functions for neural network training were studied~(see~\cite{fluctuations}).
It was conjectured that fluctuations are the consequence of inaccuracy accumulation~(see~\cite{spirals}).
Experiments~\cite{fluctuations,spirals} were reproduced for this paper with
the same datasets and precision estimation.
Here only algorithms on CPU are used, so the estimations are fully reliable.

Let us denote the experiment setting (see~\cite{fluctuations}):
\begin{itemize}
    \item We take piecewise-linear functions.
    \item Neural network gets the function argument and is trained to approximate
        the function result.
    \item We train a neural network with a single hidden layer, ReLU activation
        and MSE loss function.
        The size of the hidden layer may not coincide with the number of linear segments of the function.
    \item Training dataset is the set of values of the function at a grid of arguments.
\end{itemize}

The experiment revealed the following:
\begin{itemize}
    \item Loss function value almost everywhere begins to fluctuate, if the training was long enough.
    \item There are two periods of damped oscillations in the neural network parameter space.
        These oscillations form a double twisted spirals for adaptive momentum optimizer.
    \item It was conjectured that the fluctuations arise 
        from inaccuracy accumulation for floating point computations.
        Double twisted spirals behavior arises with this conjecture.
\end{itemize}

The experiments were reproduced with the same datasets and hyper-parameters.
The loss function values were provided with numbers of exact mantissa bits.

        \begin{figure}
            \includegraphics[scale=0.25]{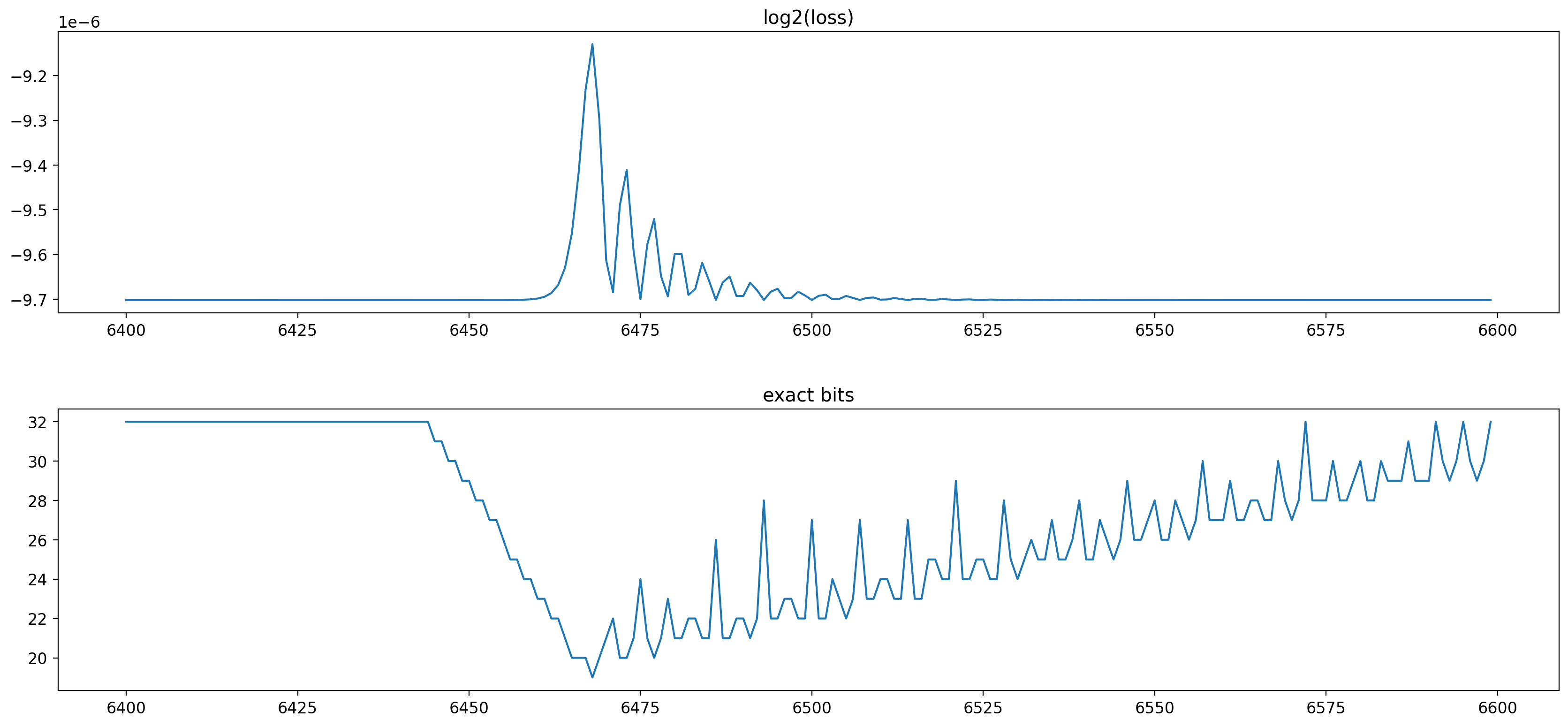}
            \caption{$3$ breaks, $7$ neurons; loss function logarithm (upper graph) and loss function mantissa exact bits (lower graph).}
            \label{fig:37}
        \end{figure}

We see the following behavior:
\begin{itemize}
    \item loss function values almost stabilize,
    \item loss function relative inexactness becomes to increase,
    \item loss function grows,
    \item loss-function and its inexactness decrease,
    \item the same repeats quasi-periodically.
\end{itemize}

\begin{figure}
    \includegraphics[scale=0.25]{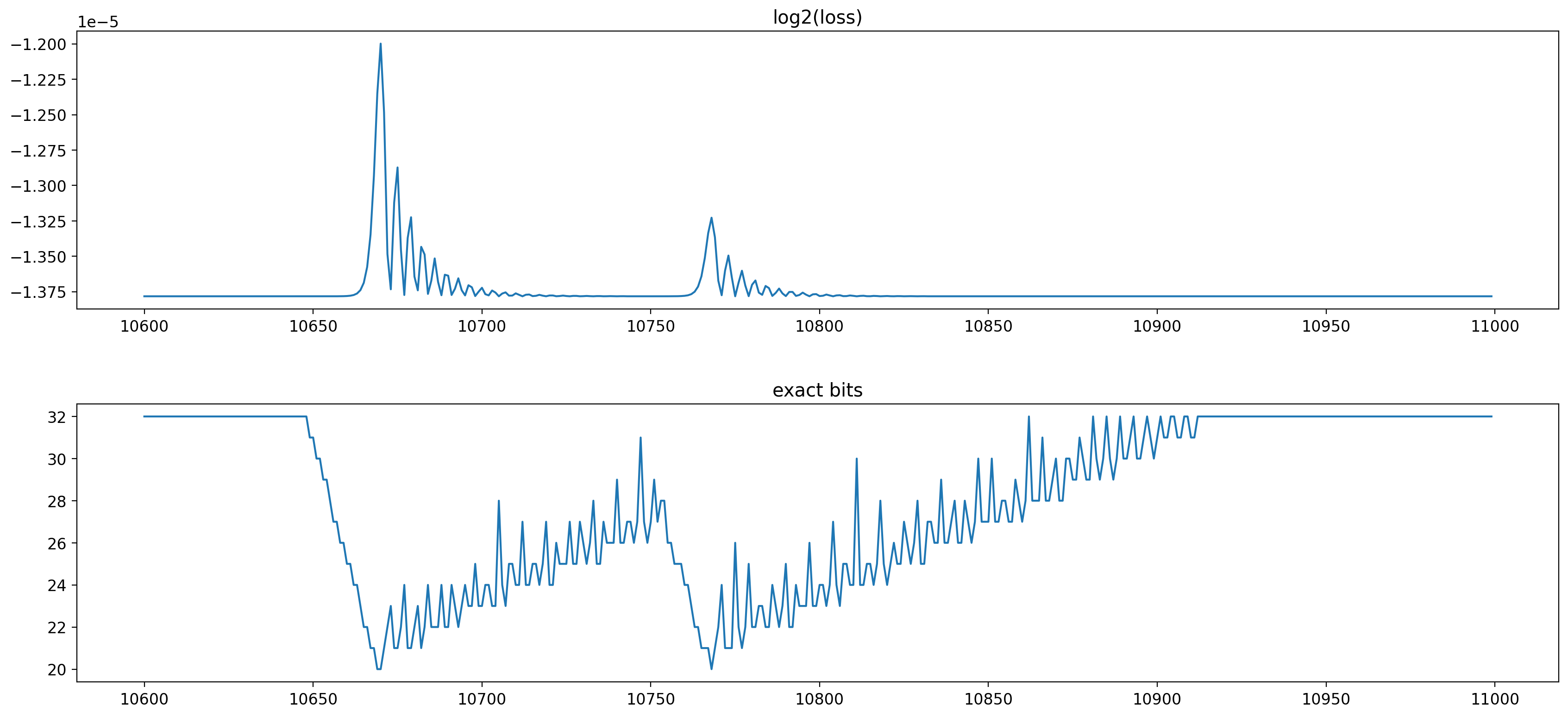}
    \caption{$3$ breaks, $8$ neurons}
\end{figure}
\begin{figure}
    \includegraphics[scale=0.25]{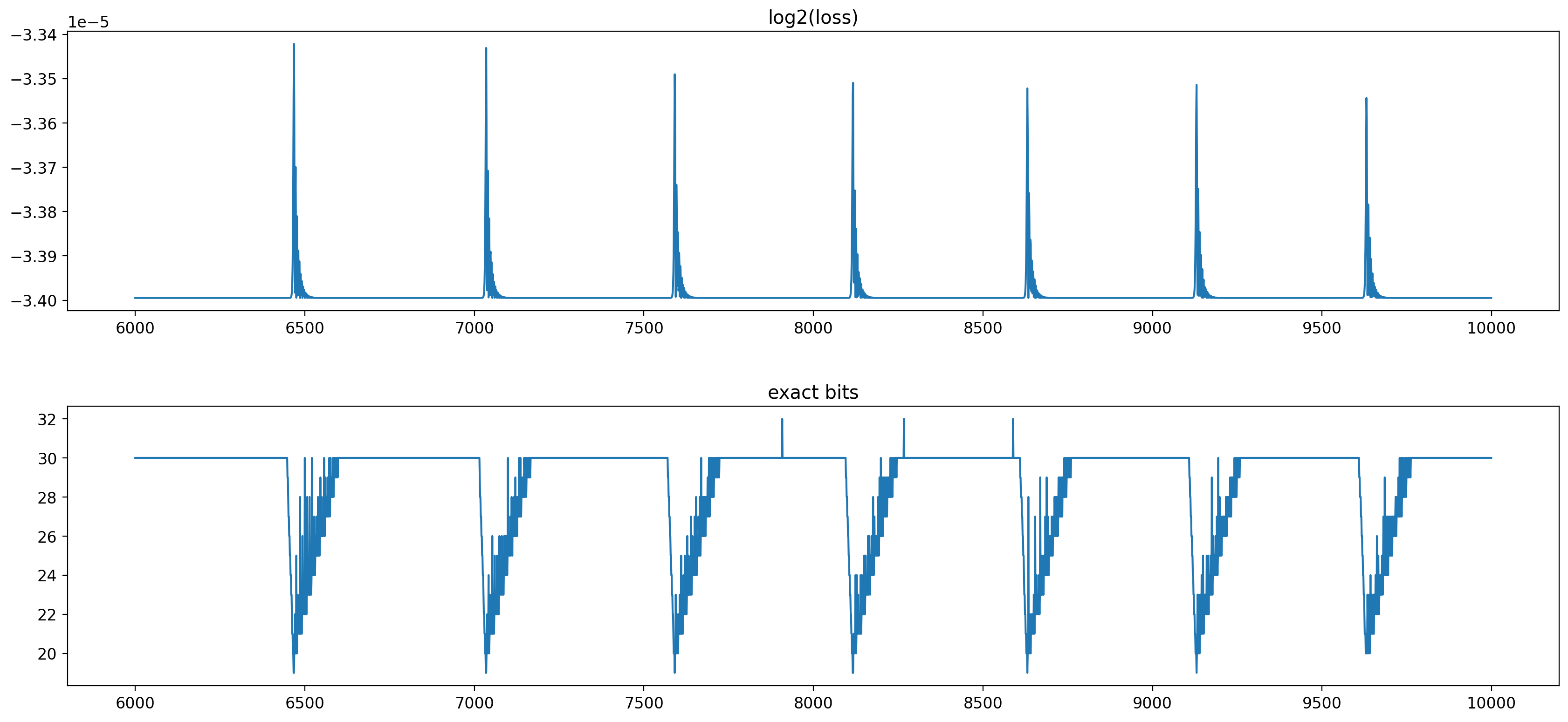}
    \caption{$3$ breaks, $3$ neurons}
\end{figure}
\begin{figure}
    \includegraphics[scale=0.25]{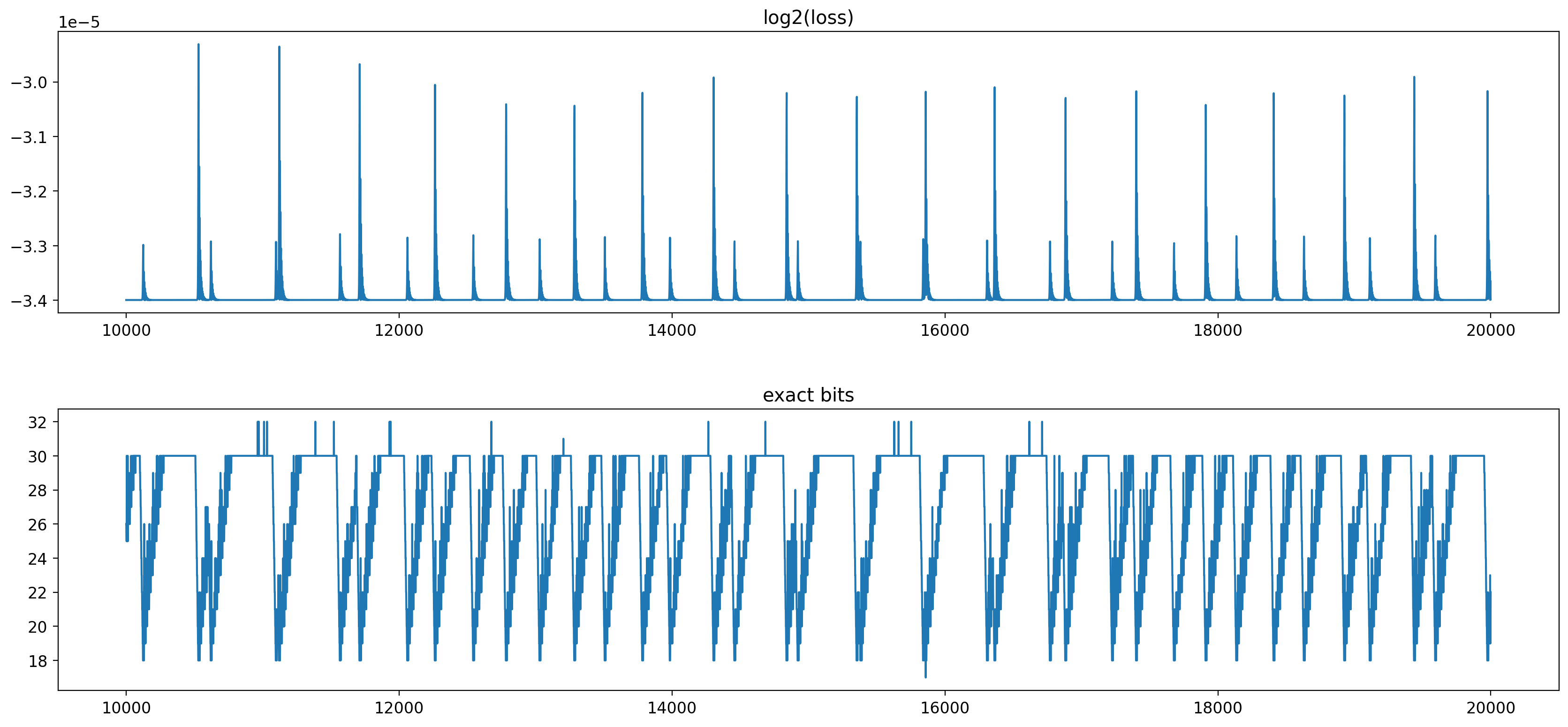}
    \caption{$3$ breaks, $5$ neurons}
\end{figure}

We see that loss function precision begins to decrease before the loss explosion,
so the loss explosion can be predicted.
Double twisted spirals also can be predicted by the loss function exact bits.
Linear precision decrease corresponds to exponential inaccuracy increase for approximately constant loss value.

\subsection{Significance loss for neural network training}

Here we conduct another numerical experiment.
This experiment setting is the following:
\begin{itemize}
    \item The neural network model consists of convolutional and linear layers and ReLU activations (it is not the LeNet model).
    \item It is trained for classification with cross-entropy loss function.
    \item The optimizer is Adam with default parameters; the batch size is~$60$.
    \item The training dataset is the standard dataset MNIST.
    \item The model achieves about $90\%$ training accuracy.
\end{itemize}

See the following training fragment:
\begin{verbatim}
batch_num=446 0.597??? 47/60
batch_num=447 0.19???? 55/60
batch_num=448 0.18???? 59/60
batch_num=449 ?.?????? 56/60
batch_num=450 0.12???? 58/60
batch_num=451 0.08???? 58/60
batch_num=452 0.1????? 55/60
batch_num=453 ?.?????? 51/60
batch_num=454 0.296??? 55/60
batch_num=455 0.24???? 54/60
batch_num=456 0.13???? 57/60
\end{verbatim}
Here batch number, the loss value and the number of correctly predicted data samples are given.
Inexact digits in the loss value are replaced with question marks.
We see that there are some batches where the loss value totally loses significance.
So, the gradients from these batches have no exact bits.
As a result, computed gradients are meaningless.
So, the training process can loose significance, although some samples may be predicted correctly,
and the loss value may decrease.

The gradients arising from batches with totally lost loss function significance
must be ignored without optimizer step.
Actually, the same holds if any entry of gradients loses significance and has no exact bits.

\section{Conclusion}

Almost any neural network accumulates floating point computational errors leading
to fluctuations of loss function.
Consequently, behavior of almost any neural network after long enough training 
becomes different from model of neural network with real-valued coefficients.
Numerical artifacts coming from floating point model (IEEE754) changes behavior from mathematical model.
So, it should be checked whether behavior is still close to mathematical model.

There are the following variants how loss of significance can break expected results.

\begin{enumerate}
    \item Loss of significance on inference can make result meaningless.
        Although the answer given by neural network may be correct, its computation
        is wrong and should not be considered.
        User should not be provided with such result.
    \item Loss of significance on training can make training meaningless.
        Although loss may decrease, optimizer step should not be performed.
        If loss function value or some gradient element has not exact bits at all,
        gradients should be zeroed without optimizer step.
    \item Loss of behavior predicted by mathematical model of neural network.
        Although this does not spoil the results and their reliability,
        it makes model less interpretable.
        Neural networks themselves are black box as a mathematical model.
        If the computational model differs from mathematical model, it makes
        resulting actual model ``double black box''.
        This breaks mathematical interpretations of models without consideration
        of floating point error arithmetic instead of real number computational analysis.
\end{enumerate}

Counting of floating point errors shows boundary of applicability of real-valued computational methods.
In the case of neural networks, long enough training at some time crosses this boundary for any network.

\section{Acknowledgments}

The author is grateful to his Kryptonite colleagues Vasily Dolmatov,
Dr. Nikita Gabdullin, and Dr. Anton Raskovalov, and Ilya Androsov for fruitful discussions of topic
and results and for assistance in testing the \texttt{xtorch} library.

\bibliographystyle{unsrt}
\bibliography{refs}

\begin{thebibliography}{10}

\bibitem{Arnold2009}
Alexander~B. Givental, Boris~A. Khesin, Jerrold~E. Marsden, Alexander~N.
  Varchenko, Victor~A. Vassiliev, Oleg~Ya. Viro, and Vladimir~M. Zakalyukin,
  editors.
\newblock {\em On functions of three variables}, pages 5--8.
\newblock Springer Berlin Heidelberg, Berlin, Heidelberg, 2009.

\bibitem{Arnold2009_2}
Alexander~B. Givental, Boris~A. Khesin, Jerrold~E. Marsden, Alexander~N.
  Varchenko, Victor~A. Vassiliev, Oleg~Ya. Viro, and Vladimir~M. Zakalyukin,
  editors.
\newblock {\em Representation of continuous functions of three variables by the
  superposition of continuous functions of two variables}, pages 47--133.
\newblock Springer Berlin Heidelberg, Berlin, Heidelberg, 2009.

\bibitem{10.1007/978-1-4612-2856-1_21}
Tianping Chen, Hong Chen, and Ruey-wen Liu.
\newblock A constructive proof and an extension of cybenko's approximation
  theorem.
\newblock In Connie Page and Raoul LePage, editors, {\em Computing Science and
  Statistics}, pages 163--168, New York, NY, 1992. Springer New York.

\bibitem{Cybenko1989ApproximationBS}
George~V. Cybenko.
\newblock Approximation by superpositions of a sigmoidal function.
\newblock {\em Mathematics of Control, Signals and Systems}, 2:303--314, 1989.

\bibitem{Funahashi1989OnTA}
Ken ichi Funahashi.
\newblock On the approximate realization of continuous mappings by neural
  networks.
\newblock {\em Neural Networks}, 2:183--192, 1989.

\bibitem{Kidger2019UniversalAW}
Patrick Kidger and Terry Lyons.
\newblock Universal approximation with deep narrow networks.
\newblock {\em ArXiv}, abs/1905.08539, 2019.

\bibitem{Geuchen2023UniversalAW}
Paul Geuchen, Thomas Jahn, and Hannes Matt.
\newblock Universal approximation with complex-valued deep narrow neural
  networks.
\newblock {\em ArXiv}, abs/2305.16910, 2023.

\bibitem{baker1998universal}
Mark~R Baker and Rajendra~B Patil.
\newblock Universal approximation theorem for interval neural networks.
\newblock {\em Reliable Computing}, 4(3):235--239, 1998.

\bibitem{kratsios2022universal}
Anastasis Kratsios and L{\'e}onie Papon.
\newblock Universal approximation theorems for differentiable geometric deep
  learning.
\newblock {\em Journal of Machine Learning Research}, 23(196):1--73, 2022.

\bibitem{voigtlaender2023universal}
Felix Voigtlaender.
\newblock The universal approximation theorem for complex-valued neural
  networks.
\newblock {\em Applied and computational harmonic analysis}, 64:33--61, 2023.

\bibitem{qi2025universal}
Qian Qi.
\newblock Universal approximation theorem of deep q-networks.
\newblock {\em arXiv preprint arXiv:2505.02288}, 2025.

\bibitem{nguyen2016universal}
Hien~D Nguyen, Luke~R Lloyd-Jones, and Geoffrey~J McLachlan.
\newblock A universal approximation theorem for mixture-of-experts models.
\newblock {\em Neural computation}, 28(12):2585--2593, 2016.

\bibitem{gonon2025universal}
Lukas Gonon and Antoine Jacquier.
\newblock Universal approximation theorem and error bounds for quantum neural
  networks and quantum reservoirs.
\newblock {\em IEEE Transactions on Neural Networks and Learning Systems},
  2025.

\bibitem{yu2021arbitrary}
Annan Yu, Chlo{\'e} Becquey, Diana Halikias, Matthew~Esmaili Mallory, and Alex
  Townsend.
\newblock Arbitrary-depth universal approximation theorems for operator neural
  networks.
\newblock {\em arXiv preprint arXiv:2109.11354}, 2021.

\bibitem{Sonoda2024ConstructiveUA}
Sho Sonoda, Yuka Hashimoto, Isao Ishikawa, and Masahiro Ikeda.
\newblock Constructive universal approximation theorems for deep
  joint-equivariant networks by schur's lemma.
\newblock {\em ArXiv}, abs/2405.13682, 2024.

\bibitem{song2017complexity}
Le~Song, Santosh Vempala, John Wilmes, and Bo~Xie.
\newblock On the complexity of learning neural networks.
\newblock {\em Advances in neural information processing systems}, 30, 2017.

\bibitem{kon2000information}
Mark~A Kon and Leszek Plaskota.
\newblock Information complexity of neural networks.
\newblock {\em Neural Networks}, 13(3):365--375, 2000.

\bibitem{liu2024kan}
Ziming Liu, Yixuan Wang, Sachin Vaidya, Fabian Ruehle, James Halverson, Marin
  Solja{\v{c}}i{\'c}, Thomas~Y Hou, and Max Tegmark.
\newblock Kan: Kolmogorov-arnold networks.
\newblock {\em arXiv preprint arXiv:2404.19756}, 2024.

\bibitem{somvanshi2024survey}
Shriyank Somvanshi, Syed~Aaqib Javed, Md~Monzurul Islam, Diwas Pandit, and
  Subasish Das.
\newblock A survey on kolmogorov-arnold network.
\newblock {\em ACM Computing Surveys}, 2024.

\bibitem{ji2024comprehensive}
Tianrui Ji, Yuntian Hou, and Di~Zhang.
\newblock A comprehensive survey on kolmogorov arnold networks (kan).
\newblock {\em arXiv preprint arXiv:2407.11075}, 2024.

\bibitem{vaca2024kolmogorov}
Cristian~J Vaca-Rubio, Luis Blanco, Roberto Pereira, and M{\`a}rius Caus.
\newblock Kolmogorov-arnold networks (kans) for time series analysis.
\newblock {\em arXiv preprint arXiv:2405.08790}, 2024.

\bibitem{kiamari2024gkan}
Mehrdad Kiamari, Mohammad Kiamari, and Bhaskar Krishnamachari.
\newblock Gkan: Graph kolmogorov-arnold networks.
\newblock {\em arXiv preprint arXiv:2406.06470}, 2024.

\bibitem{bodner2024convolutional}
Alexander~Dylan Bodner, Antonio~Santiago Tepsich, Jack~Natan Spolski, and
  Santiago Pourteau.
\newblock Convolutional kolmogorov-arnold networks.
\newblock {\em arXiv preprint arXiv:2406.13155}, 2024.

\bibitem{aghaei2024rkan}
Alireza~Afzal Aghaei.
\newblock rkan: Rational kolmogorov-arnold networks.
\newblock {\em arXiv preprint arXiv:2406.14495}, 2024.

\bibitem{cheon2024demonstrating}
Minjong Cheon.
\newblock Demonstrating the efficacy of kolmogorov-arnold networks in vision
  tasks.
\newblock {\em arXiv preprint arXiv:2406.14916}, 2024.

\bibitem{bresson2024kagnns}
Roman Bresson, Giannis Nikolentzos, George Panagopoulos, Michail
  Chatzianastasis, Jun Pang, and Michalis Vazirgiannis.
\newblock Kagnns: Kolmogorov-arnold networks meet graph learning.
\newblock {\em arXiv preprint arXiv:2406.18380}, 2024.

\bibitem{koenig2024kan}
Benjamin~C Koenig, Suyong Kim, and Sili Deng.
\newblock Kan-odes: Kolmogorov--arnold network ordinary differential equations
  for learning dynamical systems and hidden physics.
\newblock {\em Computer Methods in Applied Mechanics and Engineering},
  432:117397, 2024.

\bibitem{shen2025reduced}
Haoran Shen, Chen Zeng, Jiahui Wang, and Qiao Wang.
\newblock Reduced effectiveness of kolmogorov-arnold networks on functions with
  noise.
\newblock In {\em ICASSP 2025-2025 IEEE International Conference on Acoustics,
  Speech and Signal Processing (ICASSP)}, pages 1--5. IEEE, 2025.

\bibitem{hwang2025floating}
Geonho Hwang, Yeachan Park, Wonyeol Lee, and Sejun Park.
\newblock Floating-point neural networks can represent almost all
  floating-point functions.
\newblock In {\em Forty-second International Conference on Machine Learning},
  2025.

\bibitem{vietrov2024methodological}
Olha~Trofymenko Vietrov, Vira Trofymenko, and Maxim Hryhorov.
\newblock Methodological aspects of studying the accuracy of computer
  calculations in applied problems.
\newblock 2024.

\bibitem{xnumpy_repo}
Igor~V. Netay.
\newblock Xnumpy, 2023.

\bibitem{romanov2021analysis}
Aleksandr~Yu Romanov, Alexander~L Stempkovsky, Ilia~V Lariushkin, Georgy~E
  Novoselov, Roman~A Solovyev, Vladimir~A Starykh, Irina~I Romanova, Dmitry~V
  Telpukhov, and Ilya~A Mkrtchan.
\newblock Analysis of posit and bfloat arithmetic of real numbers for machine
  learning.
\newblock {\em IEEE Access}, 9:82318--82324, 2021.

\bibitem{agrawal2019dlfloat}
Ankur Agrawal, Silvia~M Mueller, Bruce~M Fleischer, Xiao Sun, Naigang Wang,
  Jungwook Choi, and Kailash Gopalakrishnan.
\newblock Dlfloat: A 16-b floating point format designed for deep learning
  training and inference.
\newblock In {\em 2019 IEEE 26th Symposium on Computer Arithmetic (ARITH)},
  pages 92--95. IEEE, 2019.

\bibitem{trusov2025training}
AV~Trusov.
\newblock Training 4.6-bit convolutional neural networks with a hardtanh
  activation function.
\newblock {\em Pattern Recognition and Image Analysis}, 35(1):44--64, 2025.

\bibitem{guo2025pt}
Yufei Guo, Zecheng Hao, Jiahang Shao, Jie Zhou, Xiaode Liu, Xin Tong, Yuhan
  Zhang, Yuanpei Chen, Weihang Peng, and Zhe Ma.
\newblock Pt-bitnet: Scaling up the 1-bit large language model with
  post-training quantization.
\newblock {\em Neural Networks}, page 107855, 2025.

\bibitem{saarinen1993ill}
Sirpa Saarinen, Randall Bramley, and George Cybenko.
\newblock Ill-conditioning in neural network training problems.
\newblock {\em SIAM Journal on Scientific Computing}, 14(3):693--714, 1993.

\bibitem{sinha2018neural}
Abhishek Sinha, Mayank Singh, and Balaji Krishnamurthy.
\newblock Neural networks in an adversarial setting and ill-conditioned weight
  space.
\newblock In {\em Joint European Conference on Machine Learning and Knowledge
  Discovery in Databases}, pages 177--190. Springer, 2018.

\bibitem{krishnamurthyneural}
Balaji Krishnamurthy.
\newblock Neural networks in an adversarial setting and ill-conditioned weight
  space.
\newblock In {\em Workshop Proceedings}, page~18.

\bibitem{van2002solving}
Patrick Van Der~Smagt and Gerd Hirzinger.
\newblock Solving the ill-conditioning in neural network learning.
\newblock In {\em Neural networks: tricks of the trade}, pages 193--206.
  Springer, 2002.

\bibitem{cao2025analysis}
Wenbo Cao and Weiwei Zhang.
\newblock An analysis and solution of ill-conditioning in physics-informed
  neural networks.
\newblock {\em Journal of Computational Physics}, 520:113494, 2025.

\bibitem{burger2000training}
Martin Burger and Heinz~W Engl.
\newblock Training neural networks with noisy data as an ill-posed problem.
\newblock {\em Advances in Computational Mathematics}, 13(4):335--354, 2000.

\bibitem{li2005network}
Yuanqing Li and Jun Wang.
\newblock A network model for blind source extraction in various
  ill-conditioned cases.
\newblock {\em Neural networks}, 18(10):1348--1356, 2005.

\bibitem{xnumpy}
Igor~V Netay.
\newblock Algorithms and data structures for numerical computations with
  automatic precision estimation.
\newblock {\em arXiv preprint arXiv:2403.16660}, 2024.

\bibitem{fluctuations}
Igor~V. Netay.
\newblock Influence of digital fluctuations on behavior of neural networks.
\newblock {\em Indian Journal of Artificial Intelligence and Neural Networking
  (IJAINN)}, 3:1--7, December 2022.

\bibitem{spirals}
Igor~V Netay.
\newblock Geometrical structures of digital fluctuations in parameter space of
  neural networks trained with adaptive momentum optimization.
\newblock {\em arXiv preprint arXiv:2408.12273}, 2024.

\end{thebibliography}

\end{document}